\documentclass[aps,nofootinbib,floatfix,twocolumn]{revtex4}
\usepackage{graphicx}

\begin{document}

\title{Slow roll in simple non-canonical inflation}

\author{Gabriela Barenboim} \email{Gabriela.Barenboim@uv.es}
\affiliation{Departament de F\'isica Te\`orica, Universitat de
Val\`encia, Carrer Dr. Moliner 50, E-46100 Burjassot (Val\`encia), Spain}
\author{William H.\ Kinney} \email{whkinney@buffalo.edu}
\affiliation{Dept. of Physics, University at Buffalo,
        the State University of New York, Buffalo, NY 14260-1500}
\date{\today}

\begin{abstract}
We consider inflation using a class of non-canonical Lagrangians for which the modification to the kinetic term depends on the field, but not its derivatives. We generalize the standard Hubble slow roll expansion to the non-canonical case and derive expressions for observables in terms of the generalized slow roll parameters. We apply the general results to the illustrative case of ``Slinky'' inflation, which has a simple, exactly solvable, non-canonical representation.  However, when transformed into a canonical basis, Slinky inflation consists of a field oscillating on a multi-valued potential.  We calculate the power spectrum of curvature perturbations for Slinky inflation directly in the non-canonical basis, and show that the spectrum is approximately a power law on large scales, with a ``blue'' power spectrum. On small scales, the power spectrum exhibits strong oscillatory behavior. This is an example of a model in which the widely used solution of Garriga and Mukhanov gives the wrong answer for the power spectrum. 
\end{abstract}


\maketitle

\section{Introduction}
\label{sec:introduction}

There is no doubt at present that modern cosmology leans on a consistent
theoretical framework which agrees {\it quantitatively} with data. 
Based on general relativity and the Big Bang theory, it can describe with
amazing precision the evolution of the Universe from the first fraction of a second forward. 
Nevertheless, such an impressive framework falls short of explaining
the flatness and homogeneity of space, let alone the origin of matter
and structures we observe in the universe today. As a result,  no 
self-respecting theory of the Universe can be considered  complete without a solution to these problems, the most successful of which is inflation \cite{Guth:1980zm}.

However, inflation is far from rising to the level of a theory. Inflation is just a general term for models of the very early universe which involve a  period of exponential expansion, blowing up an extremely small region to one equivalent to the current horizon size in a fraction of a second. In fact there are a bewildering variety of different models to realize  inflation. In most of them however, inflation is implemented through a single scalar field whose equations of motion are  are solved within the slow roll  approximation \cite{Albrecht:1982wi,Steinhardt:1984jj,Linde:1981mu}.

While it is clearly possible to generate a variety of interesting
inflationary models from the simplest Lagrangians, {\it i.e.} those involving a minimally coupled inflaton with a canonical kinetic term, it is a limited approach. Models with non-canonical kinetic terms are one possible way to move beyond the simplest scalar field Lagrangians. Non-canonical inflation is certainly not a new idea, having first been studied in a general sense as ``k-inflation'' \cite{Armendariz-Picon:1999rj,Garriga:1999vw}. Many particular examples of non-canonical inflation arise in the context of string-inspired inflation models such as the KKLMMT scenario \cite{Kachru:2003sx} or DBI inflation \cite{Silverstein:2003hf,Alishahiha:2004eh}. In this paper we consider a restricted class of k-inflation models for which the modification to the scalar kinetic term depends on the field but {\em not} on its derivatives. This might at first appear to be a trivial modification, since in such a case one can always transform the non-canonical scalar into a canonical one via a field redefinition. However, it is reasonable to expect that some models which appear complex or highly contrived when viewed in a canonical basis may have a simple description in an equivalent non-canonical representation. Therefore, it is useful to develop a set of tools for analyzing slow roll inflation directly in terms of a non-canonical field, without resorting to field redefinitions. 

The paper is organized as follows: In Section \ref{sec:ncinflation}, we generalize the standard slow roll expansion to the case of simple non-canonical Lagrangians, and derive expressions for the scalar and tensor power spectra generated by a non-canonical inflaton. In Section \ref{sec:slinky} we introduce the example of ``Slinky'' inflation \cite{Barenboim:2005np,Barenboim:2006rx}, which is an exactly solvable non-canonical inflation model with a highly nontrivial canonical equivalent which consists of a field oscillating on a {\em multi-valued} potential. In Section \ref{sec:perturbations} we calculate the curvature power spectrum for Slinky inflation, which displays both power-law and oscillatory regions. One very interesting feature of Slinky inflation is that the solution of Garriga and Mukhanov \cite{Garriga:1999vw}, widely used in stringy model building, can be seen to give the wrong answer. Section \ref{sec:conclusions} presents discussion and conclusions. 

\section{Inflation from a non-canonical Lagrangian}
\label{sec:ncinflation}

In this section, we derive general expressions for inflationary parameters from a Lagrangian of the form
\begin{equation}
{\cal L} = {1 \over 2} F\left(\theta\right) g^{\mu \nu} \partial_\mu \theta \partial_\nu \theta - V\left(\theta\right). 
\end{equation}
A general k-inflation model may also include a function of the field derivatives $\partial_\mu \theta$ in the kinetic term, but we concentrate on this simpler case and show that it can exhibit surprisingly complex behavior.
The Euler-Lagrange equation for an arbitrary metric $g_{\mu \nu}$ corresponding to this Lagrangian is:
\begin{eqnarray}
&&{1 \over \sqrt{-g}} \partial_\nu \left[ \sqrt{-g} F\left(\theta\right) g^{\mu \nu} \partial_\mu \theta\right] + V'\left(\theta\right)\cr
&& - {1 \over 2} g^{\mu \nu} \partial_\mu \theta \partial_\nu \theta F'\left(\theta\right) = 0.
\end{eqnarray}
Specializing to the case of a Friedmann-Robertson-Walker metric, $g_{\mu \nu} = {\rm diag}\left[1,-a^2(t),-a^2(t),-a^2(t)\right]$ and the homogeneous mode of the scalar field $\nabla\theta = 0$ results in the equation of motion
\begin{equation}
\label{eq:EOM}
\ddot\theta + 3 H \dot\theta + {1 \over F\left(\theta\right)} \left[V'\left(\theta\right) + {1 \over 2} \dot\theta^2 F'\left(\theta\right)\right] = 0,
\end{equation}
where $H = (\dot a / a)$ is the Hubble parameter. We can similarly express the stress-energy of the non-canonical field as
\begin{equation}
T^{\mu \nu} = F\left(\theta\right) \partial^\mu \theta \partial^\nu \theta - {\cal L} g^{\mu \nu},
\end{equation}
which for the homogeneous mode reduces to
\begin{eqnarray}
T^{0 0} &=& {1 \over a^2} \left[{1 \over 2} F\left(\theta\right) \dot\theta^2 + V\left(\theta\right)\right]\cr
T^{i j} &=& {1 \over a^2} \left[{1 \over 2} F\left(\theta\right) \dot\theta^2 - V\left(\theta\right)\right].
\end{eqnarray}
We can therefore express the energy density and pressure of the field $\theta$ as:
\begin{eqnarray}
\rho &=& {1 \over 2} F\left(\theta\right) \dot\theta^2 + V\left(\theta\right)\cr
p &=& {1 \over 2} F\left(\theta\right) \dot\theta^2 - V\left(\theta\right),
\end{eqnarray}
so that the speed of sound in the scalar field fluid is  \cite{Garriga:1999vw}
\begin{equation}
c_{\rm S}^2 \equiv {\partial p / \partial \dot\theta \over \partial \rho / \partial \dot\theta} = 1.
\end{equation}
From these expressions it is straightforward to show that the continuity equation
\begin{equation}
\dot\rho + 3 H \left(\rho + p\right) = 0
\end{equation}
is equivalent to the equation of motion (\ref{eq:EOM}) for the scalar field.  The Einstein Field Equation for the Hubble parameter is then
\begin{equation}
\label{eq:hubbleparam}
H^2 = {8 \pi \over 3 m_{\rm Pl}^2} \left[{1 \over 2} F\left(\theta\right) \dot\theta^2 + V\left(\theta\right)\right].
\end{equation}
Together with Eq. (\ref{eq:EOM}), this forms a complete set of equations for the evolution of the universe. 

It is convenient to re-write the equations (\ref{eq:EOM}) and (\ref{eq:hubbleparam}) in the useful Hamilton-Jacobi form \cite{grishchuk:88,Muslimov:1990be,Salopek:1990jq,Lidsey:1995np}. Differentiating Eq. (\ref{eq:hubbleparam}) with respect to time gives
\begin{eqnarray}
2 H H'\left(\theta\right) \dot\theta &=& {8 \pi \over 3 m_{\rm Pl}^2} \left[{1 \over 2} F'\left(\theta\right) \dot\theta^3 + F\left(\theta\right) \dot\theta \ddot\theta + V'\left(\theta\right) \dot\theta\right]\cr
&=& - {8 \pi \over m_{\rm Pl}^2} F\left(\theta\right) H\left(\theta\right) \dot\theta,
\end{eqnarray}
where we have used the equation of motion (\ref{eq:EOM}). Using the Friedmann Equation (\ref{eq:hubbleparam}) we then have the Hamilton-Jacobi Equations
\begin{eqnarray}
\label{eq:hamjacobi}
H'\left(\theta\right) &=& - {4 \pi \over m_{\rm Pl}^2} F\left(\theta\right) \dot\theta\cr
{8 \pi \over 3 m_{\rm Pl}^2} V\left(\theta\right) &=& H^2\left(\theta\right) \left[1 - {m_{\rm Pl}^2 \over 12 \pi F\left(\theta\right)} \left({H'\left(\theta\right) \over H\left(\theta\right)}\right)^2\right].
\end{eqnarray}
These equation are completely equivalent to the original equations (\ref{eq:EOM}) and (\ref{eq:hubbleparam}), but use the field value as a clock instead of the coordinate time, which is consistent as long as the field evolution is monotonic. We can then express the second equation above in terms of the slow roll parameter $\epsilon$, defined as
\begin{equation}
\label{eq:defepsilon}
\epsilon\left(\theta\right) \equiv {m_{\rm Pl}^2 \over 4 \pi} \left({1 \over F\left(\theta\right)}\right) \left({H'\left(\theta\right) \over H\left(\theta\right)}\right)^2,
\end{equation}
so that
\begin{equation}
{8 \pi \over 3 m_{\rm Pl}^2} V\left(\theta\right) = H^2\left(\theta\right) \left[1 - {1 \over 3} \epsilon\left(\theta\right)\right].
\end{equation}
This is exactly the same form as the Hamilton-Jacobi equation for a canonical scalar field; all of the dependence on the non-canonical function $F\left(\theta\right)$ has been absorbed into the expression (\ref{eq:defepsilon}) for the slow-roll parameter $\epsilon$. We can verify that $\epsilon$ has its usual interpretation in terms of the equation of state by writing
\begin{eqnarray}
\rho + 3 p &=& 2 F\left(\theta\right) \dot\theta^2 - 2 V\left(\theta\right)\cr
&=& - {3 m_{\rm Pl}^2 \over 4 \pi} H^2\left(\theta\right) \left[1 - \epsilon\left(\theta\right)\right].
\end{eqnarray}
The equation of state of the non-canonical field $\theta$ is then
\begin{equation}
p = \rho \left[1 - {1 \over 3} \epsilon\left(\theta\right)\right],
\end{equation}
and the condition for inflation $p < - \rho / 3$ is then, as for a canonical field, $\epsilon < 1$. The scale factor evolves as $a \propto e^{-N}$, where the number of e-folds $N$ is given by
\begin{equation}
\label{eq:numefolds}
N = - \int{H dt} = - \int{H \over \dot\theta} d\theta = {2 \sqrt{\pi} \over m_{\rm Pl}} \int{\sqrt{F\left(\theta\right) \over \epsilon\left(\theta\right)} d\theta}.
\end{equation}
Note that $N$ is defined to be the number of e-folds before the end of inflation, so that it is large at early times and decreases as inflation progresses. 
It is straightforward to verify that the lowest-order flow relation
\begin{equation}
\epsilon = {1 \over H} {d H \over d N}
\end{equation}
holds in the non-canonical case. We can define higher-order slow roll parameters in terms of the flow relations \cite{Kinney:2002qn},
\begin{equation}
\eta = \epsilon + {1 \over 2 \epsilon} \left({d \epsilon \over d N}\right) = {m_{\rm Pl}^2 \over 4 \pi} {\left(H'\left(\theta\right) / \sqrt{F\left(\theta\right)}\right)' \over \sqrt{F\left(\theta\right)} H\left(\theta\right)},
\end{equation}
\begin{eqnarray}
\xi^2 &=& \epsilon \eta + {d \eta \over d N}\cr
&=& {m_{\rm Pl}^4 \over 16 \pi^2} {\left(H' / \sqrt{F}\right) \left[\left(H' / \sqrt{F}\right)' / \sqrt{F}\right]' \over \sqrt{F} H^2},
\end{eqnarray}
and similarly for higher-order parameters. In fact it is clear that any expression involving a differential of a canonical field $d \phi$ can be translated to the non-canonical case by $d \phi \rightarrow \sqrt{F} d\theta$. Especially important is the amplitude of the curvature power spectrum,
\begin{equation}
P_{\cal R}^{1/2} = {\delta N \over \delta \phi} \delta\phi \rightarrow {\delta N \over \delta\theta} \delta\theta = {2 \sqrt{\pi} \over m_{\rm Pl}} \sqrt{F \over \epsilon} \delta \theta.
\end{equation}
Since the amplitude of the fluctuation in a canonical scalar is given by the Hubble parameter,
\begin{equation}
\delta\phi = {H \over 2 \pi},
\end{equation}
the fluctuation amplitude of the non-canonical field $\theta$ will be given by
\begin{equation}
\delta\theta = {H \over 2 \pi \sqrt{F}},
\end{equation}
and the curvature power spectrum is given by the usual expression in terms of $\epsilon$,
\begin{equation}
\label{eq:scalarps}
P_{\cal R}^{1/2} = {H \over m_{\rm Pl} \sqrt{\pi \epsilon}}.
\end{equation}
Using 
\begin{equation}
{d \over d \ln{k}} = {1 \over \epsilon - 1} {d \over d N},
\end{equation}
we see that the usual expression for the scalar spectral index applies in the non-canonical case,\footnote{An equivalent expression for the spectral index in terms of alternately defined slow roll parameters was derived in Ref. \cite{Chung:2003iu}.}
\begin{equation}
n - 1 \equiv {d \ln{P_{\cal R}} \over d\ln{k}} = 2 \eta - 4 \epsilon.
\end{equation}

This calculation may appear to be trivial, since for $F$ dependent only on $\theta$ and not its derivatives $\partial_\mu \theta$, we are free to transform the problem into a canonical basis
\begin{equation}
\label{eq:genericcanontrans}
\phi = \int{\sqrt{F} d\theta}
\end{equation}
and then use the usual canonical expressions for calculating the slow roll parameters and the inflationary observables. In the next section, we consider the case of ``Slinky'' inflation \cite{Barenboim:2005np,Barenboim:2006rx}, for which the transformation (\ref{eq:genericcanontrans}) is multi-valued, and therefore the associated canonical field evolution is nontrivial. 

\section{Slinky Inflation}
\label{sec:slinky}

\begin{figure}
\centerline{\includegraphics[width=3.5in]{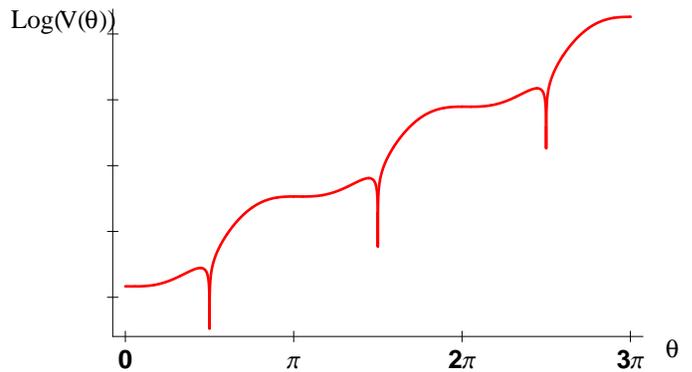}}
\caption{\label{fig:slinkypotential} The potential for Slinky inflation in terms of the non-canonical field $\theta$. Inflation takes place near the ``flat'' regions around $\theta = 0, \pm\pi, \ldots$.}
\end{figure}

A ``Slinky'' is a large spring with a very weak spring constant which ``walks''
down a staircase, producing in time-lapse a pattern resembling the potential
describing the Slinky model, {\it i.e.} 
\begin{eqnarray}
V(\theta ) &=& V_0\, {\rm cos}^2\,\theta\, {\rm exp}\left[{3\over b}\left(
2\theta - {\rm sin}\,2\theta \right) \right] \; ,
\label{eqn:ourpot} 
\end{eqnarray}
where $V_0$ is the dark energy
density observed today and whose kinetic function is chosen to be
\begin{eqnarray}\label{eqn:ourkin}
F(\theta ) &=& {3 m_{\rm Pl}^2\over\pi b^2}\, {\rm sin}^2\,\theta\: . 
\end{eqnarray}
where $b$ is an arbitrary real parameter which governs the number
of steps of the potential. Figure (\ref{fig:slinkypotential}) shows the Slinky potential. 

This apparently harmless combination of potential and kinetic terms
produces not only an unusual cosmological history but
also breaks down the possibility of switching between the canonical
and non-canonical descriptions, as the tool to do so, $F(\theta )$, 
periodically goes to zero, which results in
divergences in the slow-roll parameters. The model is characterized by a periodic equation of state,
\begin{eqnarray}
w(a) = -{\rm cos}\,2\theta (a) \; .
\end{eqnarray}
In this way, the same field can serve the role both of the inflaton in the early universe and the quintessence field, which drives the accelerated expansion in the current universe. The model also assumes that the inflaton/quintessence field has weak perturbative
couplings with matter and radiation. This introduces two 
additional degrees of freedom, so that Slinky has only three
adjustable parameters, which are chosen such that the 
 $\Omega_r/\Omega_{\Lambda}$ and  $\Omega_m/\Omega_{\Lambda}$
come out to their measured values at $a=1$ in a flat Universe. The temperature history of the model is quite nonstandard as well,
in that the coupling between radiation and the non canonical scalar field
force them track each other. Here we ignore the late-time behavior of the Slinky field and concentrate on the period of inflation in the early universe.

\begin{figure}
\centerline{\includegraphics[width=3.5in]{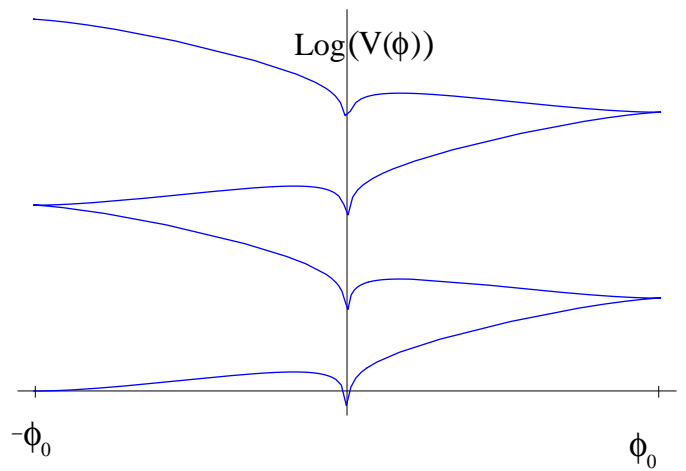}}
\caption{\label{fig:canonicalpotential} The canonical equivalent of the Slinky potential. In contrast to the monotonic non-canonical field $\theta$, the corresponding canonical field $\phi$ oscillates on a multi-valued potential, with inflation happening near the stationary points where $\dot\phi = 0$.}
\end{figure}

Despite the apparently complicated form of the potential (\ref{eqn:ourpot}), the Hamilton-Jacobi Equations (\ref{eq:hamjacobi}) can be solved exactly:
\begin{eqnarray}
\label{eq:HJsolution}
&&H\left(\theta\right) = H_0 \exp\left[\left({3 \over 2 b}\right) \left(2 \theta - \sin{2 \theta}\right)\right]\cr
&&\dot\theta = - {b \over 2} H\left(\theta\right).
\end{eqnarray}
We then have exact expressions for the slow roll parameters,
\begin{eqnarray}
\label{eq:slinkysrparams}
\epsilon\left(\theta\right) &&= 3 \sin^2{\theta}\cr
\eta\left(\theta\right) &&= 3 \sin^2{\theta} + {b \over 2 \tan{\theta}}.
\end{eqnarray}
The equation of state is oscillatory, with $\epsilon$ varying from $\epsilon = 3$ ($p = \rho$) to $\epsilon = 0$ ($p = -\rho$) periodically while the field value $\theta$ and the Hubble parameter $H$ both decrease monotonically. This behavior is not possible for a canonical field, since $\epsilon = 0$ implies $\dot\phi = 0$ for a canonical field $\phi$. Taking the interval around $\theta = 0$ for definiteness, inflation occurs for $\epsilon < 1$, or
\begin{equation}
- \sin^{-1}\left({1 \over \sqrt{3}}\right) < \theta < \sin^{-1}\left({1 \over \sqrt{3}}\right).
\end{equation}
The number of e-folds of inflation (\ref{eq:numefolds}) during this period is
\begin{equation}
\label{eq:slinkyefolds}
N = {2 \over b} \int_{\theta_f}^{\theta_i} d \theta = {4 \over b} \sin^{-1}\left({1 \over \sqrt{3}}\right) = {2.46 \over b},
\end{equation}
so that $N > 60$ requires $b < 0.04$.\footnote{Slinky inflation naturally allows for a large amount of entropy production after inflation, so that the  number of e-folds required to solve the horizon problem can in principle be much smaller than 60. For the purposes of this paper, we will assume scales of order the CMB quadrupole exited the horizon around $N = 60$.} Note that the Slinky model is strongly non-slow roll near $\theta = 0, \pm \pi, \pm 2 \pi, \ldots$, since the parameter $\eta\left(\theta\right)$ diverges at these points even as the equation of state approaches the de Sitter limit, $\epsilon = 0$. The expression (\ref{eq:scalarps}) formally diverges in the $\epsilon = 0$ limit as well, but (as we show below) this divergence is nonphysical. The physical meaning of the divergence in the parameters can be understood by transforming to a canonical basis. We can transform to a canonical field $\phi$ by taking
\begin{equation}
d \phi = \sqrt{F} d\theta = \sqrt{3 m_{\rm Pl}^2 \over \pi b^2} \sin{\theta} d\theta,
\end{equation}
so that
\begin{equation}
\label{eq:canontrans}
\phi = - \phi_0 \cos{\theta},
\end{equation}
with
\begin{equation} 
\phi_0 = m_{\rm Pl} \sqrt{3 \over \pi b^2}.
\end{equation}
From Eqs. (\ref{eq:slinkysrparams}) and (\ref{eq:canontrans}), we also have an exact expression for $\epsilon$ in terms of the canonical field $\phi$:
\begin{equation}
\label{eq:epsiloncanon}
\epsilon\left(\phi\right) = 3 \left[1 - \left({\phi \over \phi_0}\right)^2\right],
\end{equation}
and inflation occurs for $\left(\phi / \phi_0\right)^2 > 2/3$, or
\begin{equation}
\left({\phi \over m_{\rm Pl}}\right)^2 > {2 \over \pi b^2}.
\end{equation}
The canonical expression for $\epsilon$ (\ref{eq:epsiloncanon}) can be recognized as a special case of the non-slow roll solution derived in Ref. \cite{Kinney:1997ne},
\begin{equation}
\epsilon \simeq 3 \left[1 - {V\left(\phi\right) \over V\left(\phi_0\right)}\right],
\end{equation}
where $\phi_0$ is a stationary value of the field, $\dot\phi = 0$. Near the stationary value, the potential is approximately
\begin{equation}
{V\left(\phi\right) \over V\left(\phi_0\right)} \simeq 1 - {1 \over 3} \epsilon = \left({\phi \over \phi_0}\right)^2.
\end{equation}
So we expect that the canonical equivalent of Slinky evolution is qualitatively that of a field evolving near a classical turning point on a potential that is approximately quadratic in the vicinity of the turning point. This explains the apparent divergence in the power spectrum, since the standard slow roll expressions break down for a field evolving near a turning point \cite{Seto:1999jc}. However, we can be more precise than this. Since the expresion for $\epsilon$ (\ref{eq:epsiloncanon}) is exact, we can derive a corresponding exact expression for the potential \cite{Kinney:1997ne},
\begin{equation}
V\left(\phi\right) = V\left(\phi_0\right) \exp\left[{4 \sqrt{\pi} \over m_{\rm Pl}} \int_{\phi_0}^{\phi} \sqrt{\epsilon\left(\phi\right)} d\phi\right] \left(1 - {1 \over 3} \epsilon\left(\phi\right)\right).
\end{equation}
It is straightforward to use this expression to recover the original expression for the Slinky potential by transforming the integral into the non-canonical basis:
\begin{eqnarray}
\label{eq:ncintegral}
{4 \sqrt{\pi} \over m_{\rm Pl}} \int_{\phi_0}^{\phi} \sqrt{\epsilon\left(\phi\right)} d\phi &&= {4 \sqrt{3 \pi} \over m_{\rm Pl}} \phi_0 \int_0^\theta \sin^2{\theta} d\theta\cr
&&= \left({3 \over b}\right) \left(2 \theta - \sin{2 \theta}\right),
\end{eqnarray}
so that we recover exactly the Slinky potential (\ref{eqn:ourpot}).
We can also solve the canonical equivalent of the integral (\ref{eq:ncintegral}),
\begin{eqnarray}
\label{eq:F1}
&&{4 \sqrt{\pi} \over m_{\rm Pl}} \int_{\phi_0}^{\phi} \sqrt{\epsilon\left(\phi\right)} d\phi = 4 \sqrt{3 \pi}\left({\phi_0 \over m_{\rm Pl}}\right) \int_1^x{\sqrt{1 - x^2} dx}\cr
&&= \left({3 \over 2 b}\right)\left[\left({\phi \over \phi_0}\right) \sqrt{1 - \left({\phi \over \phi_0}\right)^2} + \sin^{-1}{\left({\phi \over \phi_0}\right)}\right]\cr
&&\equiv f_1\left(\phi\right).
\end{eqnarray}
The corresponding canonical potential is then
\begin{equation}
\label{eq:canonV1}
V_1\left(\phi\right) = V_0 \left({\phi \over \phi_0}\right)^2 \exp\left[f_1\left(\phi\right)\right],
\end{equation}
where $f_1\left(\phi\right)$ is defined by Eq. (\ref{eq:F1}). 
Here we take the sign convention that $\sqrt{\epsilon}$ has the same sign as $H'\left(\phi\right)$,
\begin{equation}
\sqrt{\epsilon} = + {m_{\rm Pl} \over 2 \sqrt{\pi}} {H'\left(\phi\right) \over H\left(\phi\right)},
\end{equation}
so that $\sqrt{\epsilon} > 0$ implies $\dot\phi < 0$, and the solution (\ref{eq:canonV1}) therefore represents the potential for a field rolling from $\phi = +\phi_0$ to $\phi = - \phi_0$, where $\pm \phi_0$ are turning points in the field evolution. We can similarly solve for the evolution of the field from $\phi = -\phi_0$ to $\phi = + \phi_0$. In this case, for $\dot\phi > 0$, we must take $\sqrt{\epsilon} < 0$, and
\begin{eqnarray}
\label{eq:F2}
&&{4 \sqrt{\pi} \over m_{\rm Pl}} \int_{-\phi_0}^{\phi} \sqrt{\epsilon\left(\phi\right)} d\phi = 4 \sqrt{3 \pi}\left({\phi_0 \over m_{\rm Pl}}\right) \int_x^{-1}{\sqrt{1 - x^2} dx}\cr
&&= - \left({3 \over 2 b}\right)\left[\left({\phi \over \phi_0}\right) \sqrt{1 - \left({\phi \over \phi_0}\right)^2} - \cos^{-1}{\left({\phi \over \phi_0}\right)} + \pi\right]\cr
&&\equiv f_2\left(\phi\right).
\end{eqnarray}
The corresponding potential is:
\begin{equation}
\label{eq:canonV2}
V_2\left(\phi\right) = V_1\left(-\phi_0\right) \left({\phi \over \phi_0}\right)^2 \exp\left[f_2\left(\phi\right)\right].
\end{equation}
Note that this potential is {\em not} the same as the potential (\ref{eq:canonV1}): the field rolls to the left along one potential, but then rolls back to the right along a different potential! We have chosen the normalization of Eq. (\ref{eq:canonV2}) to ensure continuity at the turning point, $V_2\left(-\phi_0\right) = V_1\left(-\phi_0\right)$. Monotonic evolution of the non-canonical scalar $\theta$ corresponds to oscillations of the canonical scalar $\phi$ on a {\em multi-valued} potential, with each oscillation rolling down a successively lower ``ramp'', as shown in Fig. \ref{fig:canonicalpotential}. The Hamilton-Jacobi equation, valid only for monotonic field evolution, can nonetheless be exactly solved for the canonical field in a piecewise fashion. 

``Slinky'' inflation provides an example of a simple non-canonical inflation model with a highly non-trivial canonical counterpart, consisting of a scalar field oscillating on a multi-valued potential. In the next section, we discuss the generation of density perturbations in Slinky models. 

\section{The Curvature Perturbation in Slinky Inflation}
\label{sec:perturbations}

In the standard slow roll approximation, we can write the power spectrum of curvature perturbations by evaluating the mode amplitude at horizon crossing,
\begin{equation}
\label{eq:SRpowerlaw}
P_{\mathcal R} = \left. {H^2 \over m_{\rm Pl}^2 \pi \epsilon}\right\vert_{k = a H}. 
\end{equation}
This is the solution of Garriga and Mukhanov \cite{Garriga:1999vw}. In the limit of slow roll, the power spectrum is approximately a power law, with spectral index given in terms of the slow roll parameters by
\begin{equation}
\label{eq:SRn}
n - 1 = 2 \eta - 4 \epsilon = -6 \sin^2{\theta} + {b \over 2 \tan{\theta}}.
\end{equation}
For this expression to be valid, $\epsilon$ and $\eta$ must be much smaller than unity. From Eq. (\ref{eq:slinkysrparams}) this results in the conditions
\begin{equation}
{b \over 2} \ll \theta \ll {1 \over \sqrt{3}},
\end{equation}
both of which can be satisfied in the limit $b \ll 1$. For $\theta \ll 1$,
\begin{equation}
n - 1 \simeq {b \over 2 \theta},
\end{equation}
so that in slow roll, with $\theta \gg b/2$, we expect a nearly power-law power spectrum, with a ``blue'' tilt ({\it i.e.} $n > 1$). Since $\dot\theta < 0$, shorter-wavelength modes cross outside the horizon at smaller values of $\theta$, so that we expect the spectrum to run such that the blue tilt will be steeper on small scales than on large scales, {\it i.e.} a ``positive'' running. 

However, slow roll inevitably breaks down as $\theta \rightarrow 0$, and the second slow roll parameter $\eta$ becomes singular, as does the horizon-crossing expression for the power spectrum (\ref{eq:SRpowerlaw}). The singularity corresponds to the turning point of the equivalent canonical field $\phi$. It is straightforward to see that the equation for quantum fluctuations in the inflaton is also singular. In terms of the conformal time $\tau$, the equation for the gauge-invariant Mukhanov variable $u_k$ is \cite{Mukhanov:1990me}
\begin{equation}
\label{eq:basemode}
{d^2 u_k \over d \tau^2} + \left(k^2 - {1 \over z} {d^2 z \over d\tau^2}\right) u_k = 0,
\end{equation}
where 
\begin{equation}
z \equiv 2 \sqrt{\pi} \left({a \dot \phi \over H}\right) = 2 \sqrt{\pi} \left({a \sqrt{F} \dot \theta \over H}\right).
\end{equation}
Using Eq. (\ref{eq:HJsolution}), we can write $z$ as:
\begin{equation}
z = - \sqrt{3} m_{\rm Pl} a_0 e^{- 2 \theta / b} \sin{\theta},
\end{equation}
where $a_0$ is defined as the arbitrary constant of proportionality in the evolution of the scale factor
\begin{equation}
a\left(\theta\right) = a_0 e^{-2 \theta / b}.
\end{equation}
We can relate the conformal time to the field value by
\begin{equation}
{2 \over b} d \theta = (a H) d \tau,
\end{equation}
and the mode equation (\ref{eq:basemode}) can be written {\em exactly} in terms of the field $\theta$ as \cite{Barenboim:2005np}:
\begin{eqnarray}
\label{eq:exactmodeequation}
&&\left({b^2 \over 4}\right) {d^2 u_k \over d\theta^2} + \left({b \over 2}\right) \left(3 \sin^2{\theta} - 1\right) {d u_k \over d\theta} + \cr
&&\left[\left({k \over a H}\right)^2 - {1 \over 4} \left(2 - b^2 + 6 \cos{2 \theta} - {6 b \cos^3{\theta} \over \sin{\theta}}\right)\right] u_k\cr &&= 0.
\end{eqnarray}
Note that the last term is singular at $\theta = 0$. We can express the wavelength of the mode relative to the horizon size as
\begin{equation}
y \equiv {k \over a H} = y_0 \exp\left[- {1 \over b} \left(\theta - {3 \over 2} \sin{2 \theta}\right)\right],
\end{equation}
where the constant of proportionality $y_0$ is evaluated at $\theta = 0$,
\begin{equation}
\label{eq:y0}
y_0 \equiv {k \over a H}\bigg\vert_{\theta = 0}.
\end{equation}
Defining
\begin{equation}
x \equiv {2 \theta \over b},
\end{equation}
we can write the mode equation near the singular point $\theta \ll 1$ as approximately
\begin{equation}
\label{eq:smallxsol}
u_k''\left(x\right) - u_k'\left(x\right) + \left[y_0^2 e^{2 x} + {3 \over x} + {b^2 \over 4} - 2\right] u_k(x) = 0.
\end{equation}
The variable $x$ can be seen from Eq. (\ref{eq:slinkyefolds}) to be a measure of the number of e-folds of inflation, and is here evolving from positive to negative. This can be solved analytically for $b \ll 1$ and for modes much larger than the horizon, $y \simeq y_0 e^{x} \rightarrow 0$
\begin{equation}
u_k = \left\lbrace{x e^{- x} \atop e^{-x} \left[e^{3 x} - 3 x {\rm Ei}\left(3 x\right)\right]}\right\rbrace.
\end{equation}
Both solutions are regular at the singularity $x = 0$, but the second has a singular derivative. Since, for $x \ll 1$, 
\begin{equation}
z \propto x e^{-x},
\end{equation}
the first solution in Eq. (\ref{eq:smallxsol}) represents the constant mode of the power spectrum on superhorizon scales,
\begin{equation}
\left\vert{u_k \over z}\right\vert \rightarrow {\rm const.},\ y \rightarrow 0.
\end{equation}
The second solution is a decaying mode. Contact can also be made with the slow roll limit when
\begin{equation}
{3 \over x} \ll {b^2 \over 4}. 
\end{equation}
where Eq. (\ref{eq:smallxsol}) has solution
\begin{equation}
u_k \propto e^{x/2} H_{\nu}\left(y_o e^{x}\right) = \sqrt{y} H_{\nu}(y),
\end{equation}
where we have applied the Bunch-Davies boundary condition $u_k \propto e^{-i k \tau}$ when $k \gg a H$, and
\begin{equation}
\nu = {1 \over 2} \sqrt{9 - b^2} \simeq {3 \over 2} \left(1 - {b^2 \over 18}\right).
\end{equation}
This solution is valid both inside and outside the horizon, as long as slow roll is a good approximation. In the long wavelength limit, 
\begin{equation}
P_{\cal R} \propto k^3 \left\vert{u_k \over z}\right\vert^2 \propto y^{3 - 2 \nu},
\end{equation}
and the scalar spectral index is given by
\begin{equation}
n  = 4 - 2 \nu \simeq 1 + {b^2 \over 6},
\end{equation}
so that in the extreme slow roll limit, we expect to generate an almost exactly scale-invariant power spectrum, with a very tiny blue tilt. 

\begin{figure}
\centerline{\includegraphics[width=3.8in]{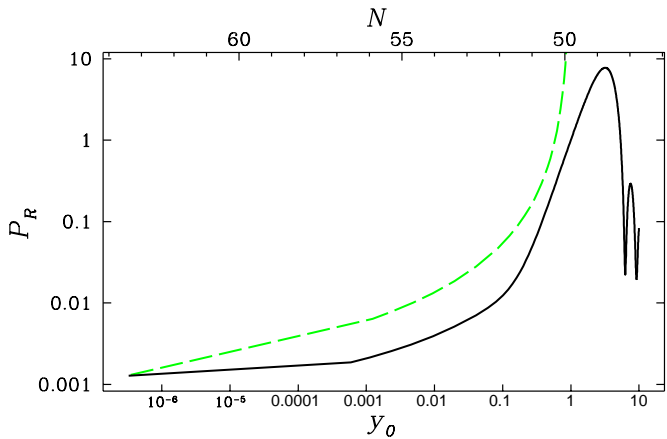}}
\caption{\label{fig:PSb0.025} The power spectrum (normalized to unity at $y_0 = 1$) for Slinky inflation plotted as a function of $\ln(y_0) \propto \ln(k)$, for $b = 0.025$. The top axis shows the number of e-folds $N$ before the end of inflation. The green (dashed) line is the slow roll solution (\ref{eq:SRpowerlaw}). (Despite its superficial resemblance to the $C_{\ell}$ spectrum of the Cosmic Microwave Background, this plot shows the primordial power spectrum!) }
\end{figure}

\begin{figure}
\centerline{\includegraphics[width=3.8in]{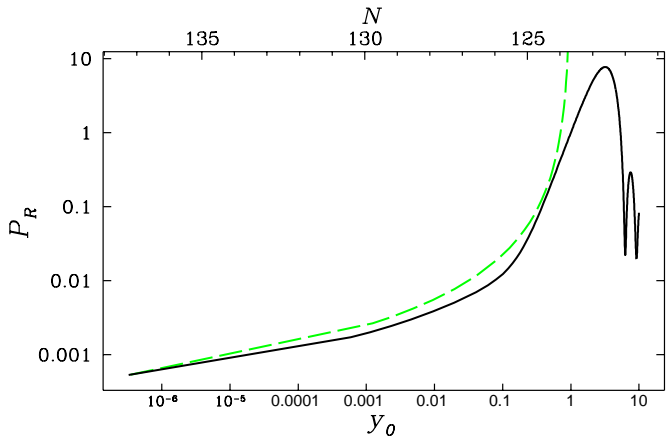}}
\caption{\label{fig:PSb0.01} The power spectrum (normalized to unity at $y_0 = 1$) for Slinky inflation plotted as a function of $\ln(y_0) \propto \ln(k)$, for $b = 0.01$. The top axis shows the number of e-folds $N$ before the end of inflation. The green (dashed) line is the slow roll solution (\ref{eq:SRpowerlaw}).}
\end{figure}

For interesting values of the model parameters, the slow roll limit is strongly violated, and we resort to numerically solving the exact mode evolution equation (\ref{eq:exactmodeequation}) mode-by-mode in $k$. For this purpose, it is useful to write Eq. (\ref{eq:exactmodeequation}) as:
\begin{eqnarray}
&&u''(x) + \left(3 \sin^2{\left(b x / 2\right)} - 1\right) u'(x) + \cr
&&\left[y_0^2 Y^2(x) - {3 \over 2} \cos{\left(b x\right)} + {3 b \cos^3{\left(b x / 2\right)} \over 2 \sin{\left(b x / 2\right)}} + {b^2 \over 4} - {1 \over 2} \right] u(x)\cr &&= 0,
\end{eqnarray}
where
\begin{equation}
Y(x) = {y(x) \over y_0} = \exp{\left[-{x \over 2} + {3 \sin{\left(b x\right)} \over 2 b}\right]}.
\end{equation}
Expressed in this form, the choice of wavenumber $k$ is mapped entirely to the choice of dimensionless parameter $y_0$ as defined by Eq. (\ref{eq:y0}). We set the boundary condition on the mode at short wavelength using the slow roll expression.
\begin{equation}
\label{eq:bc}
k^{1/2} u_k = {\sqrt{\pi} \over 2} \sqrt{- k \tau} H_\nu\left(- k \tau\right).
\end{equation}
This boundary condition corresponds to the Bunch-Davies vacuum in the ultraviolet limit, and will be accurate as long as a given mode spends a sufficiently long time inside the horizon during slow roll evolution. Since the duration of inflation is finite, this approximation will break down for modes which have wavelengths close to, or larger than, the horizon size at the begining of inflation. For these modes, the initial condition is set by the pre-inflationary evolution.\footnote{For a discussion of the effect of the pre-inflationary vacuum in a general setting, see Ref. \cite{Powell:2006yg} and references therein.} This will modify the power spectrum only at the largest scales, which we neglect in this analysis and simply use the boundary condition (\ref{eq:bc}) for all modes. Each mode is integrated from $k / (a H) = 100$ to the end of inflation. In terms of the variable $x$, this corresponds to evolution from
\begin{equation}
x_i = \ln{100 / y_0},
\end{equation}
to
\begin{equation}
x_f = - {2 \over b} \sin^{-1}\left({1 \over \sqrt{3}}\right). 
\end{equation}
In order to correctly set the boundary condition (\ref{eq:bc}) on the mode function, we must relate the conformal time $\tau$ to the field value $x$. In the slow roll limit, we have
\begin{equation}
- k \tau = {y(x) \over 1 - \epsilon}.
\end{equation}
In the non-slow roll case, this generalizes to the exact expression
\begin{equation}
d(- k \tau) = {d y \over 1 - \epsilon} = {dy(x) \over dx} {dx \over 1 - \epsilon(x)},
\end{equation}
which is straightforward to evaluate using the identity
\begin{equation}
{dy(x) \over dx} = y(x) \left[1 - \epsilon(x)\right],
\end{equation}
so that
\begin{equation}
-k \tau = \int_{x_f}^x{ y(x) dx}.
\end{equation}
We evaluate this integral numerically for each mode. 
To relate the choice of $y_0$ to a physical scale in the universe today, it is useful to calculate $N_H$, defined as the number of e-folds before the end of inflation at which a given mode crossed outside the horizon. The current CMB quadrupole then corresponds to $N_H$ of around $60$. This is straightforward, since 
\begin{equation}
e^{N_H} \equiv {a(x_f) \over a(x_H)} \simeq e^{(x_H - x_f)},
\end{equation}
where $x_H$ horizon crossing for a particular mode, $y(x_H) = y_0 e^{x_H} = 1$.  We then have
\begin{equation}
y_0 \simeq e^{-x_H},
\end{equation}
and the number of e-folds is
\begin{equation}
N_H(y_0) \simeq - \ln(y_0) - x_f.
\end{equation}
Figure \ref{fig:PSb0.025} shows the power spectrum as a function of $y0$ and $N_H$ for $b = 0.025$, which gives a total amount of inflation of around a hundred e-folds.  The power spectrum has qualitatively different behaviors for modes with $y_0 < 1$, which exit the horizon prior to the singularity at $\theta = 0$, and those with $y_0 > 1$, which exit after. This results in an approximately power-law behavior on large scales, and an oscillatory power spectrum on small scales. Note that the slow roll expression for the power spectrum, (\ref{eq:SRpowerlaw}) is a poor fit to the exact power spectrum.  The shape of the power spectrum is qualitatively similar for other choices of the parameter $b$, with power-law behavior at large scales giving way to oscillations at small scales. If we make $b$ smaller, the fit to the slow roll expression at large scales improves significantly. Figure \ref{fig:PSb0.01} shows the power spectrum for $b = 0.01$. Slow roll is a good fit to the exact result at very large scales, with $N > 125$, which are far outside the current cosmological horizon at $N = 60$. 
In any case, the WMAP3 data strongly disfavor models with a blue spectrum and negligible tensor modes \cite{Spergel:2006hy,Alabidi:2006qa,Seljak:2006bg,Kinney:2006qm,Martin:2006rs}, and perturbations from a single period of Slinky Inflation prove to be a poor fit to the real world. 
\section{Conclusions}
\label{sec:conclusions}

We have considered inflation driven by a scalar field with a Lagrangian of the form
\begin{equation}
{\cal L} = {1 \over 2} F\left(\theta\right) g^{\mu \nu} \partial_\mu \theta \partial_\nu \theta - V\left(\theta\right). 
\end{equation}
Since the modification to the kinetic term depends on the field but not its derivatives, it is possible to write an equivalent canonical Lagrangian via a field redefinition $d \phi = \sqrt{F} d\theta$. 
It is then straightforward to generalize the standard Hubble slow roll expansion to derive expressions for the flow parameters $\epsilon$, $\eta$, and so forth in terms of the field $\theta$ and the function $F\left(\theta\right)$, for example
\begin{equation}
\epsilon\left(\theta\right) = {m_{\rm Pl}^2 \over 4 \pi} \left({1 \over F\left(\theta\right)}\right) \left({H'\left(\theta\right) \over H\left(\theta\right)}\right)^2,
\end{equation}
and 
\begin{equation}
\eta = {m_{\rm Pl}^2 \over 4 \pi} {\left(H'\left(\theta\right) / \sqrt{F\left(\theta\right)}\right)' \over \sqrt{F\left(\theta\right)} H\left(\theta\right)}.
\end{equation}
In terms of these parameters, the expressions for the scalar power spectrum and the associated spectral index take on their usual forms in terms of the slow roll parameters,
\begin{equation}
P_{\cal R}^{1/2} = {H \over m_{\rm Pl} \sqrt{\pi \epsilon}},
\end{equation}
and
\begin{equation}
n - 1 \equiv {d \ln{P_{\cal R}} \over d\ln{k}} = 2 \eta - 4 \epsilon.
\end{equation}

Such a field redefinition might appear to be trivial, since one could just as easily work with the equivalent canonical field $\phi$. However, we describe a simple example, ``Slinky'' inflation, in which the transformation from the non-canonical to the canonical basis is multi-valued. In this case, a field evolving monotonically along a potential in a non-canonical representation corresponds in the canonical representation to a field oscillating on a multi-valued potential. The non-canonical representation proves much simpler, and the evolution equations for the background cosmology can be solved exactly, consisting of alternating periods of accelerating and decelerating expansion, and has been proposed as a unified model of inflation and quintessence \cite{Barenboim:2005np,Barenboim:2006rx}. In this paper, we do not consider the late-time behavior of the Slinky field, but instead concentrate on the early inflationary period of the model. We write an exact equation for the evolution of gauge-invariant curvature perturbations during inflation, which is complicated by the presence of  singularities at $\theta = 0, \pm\pi, \ldots$, which in the canonical description correspond to the extrema of the oscillating field, where the field velocity $\dot\phi = 0$. We evaluate the mode equation numerically, and find that, for modes which exit the horizon prior to the singular point, the power spectrum is an approximate power law with a ``blue'' ({\it i.e.} $n > 1$) power spectrum. Modes with shorter wavelength which are still inside the horizon at the singular point produce an oscillatory power spectrum. In the slow roll limit, the solution of Garriga and Mukhanov \cite{Garriga:1999vw}, in which the power spectrum is evaluated when a mode exits the horizon, provides a good fit to the exact power spectrum However, the solution fails in the non-slow roll regime, in particular near the singularity. As a whole, the perturbations produced in Slinky inflation are not a good fit to the current data. However, we have only considered perturbations generated during a {\em single} epoch of inflation. Since Slinky evolution is periodic, perturbations may in principle be processed through multiple periods of inflation and decelerated expansion in the early universe, and it is unclear what the effect of such evolution would be on the primordial power spectrum. 

\section*{Acknowledgments}
We thank Brian Powell for helpful conversations. WHK thanks the Institut d'Astrophysique de Paris, where part of this work was completed, for hospitality. WHK is supported in part by the National Science Foundation under grant NSF-PHY-0456777. 
GB is grateful for the support from the Spanish MEC under
Contract PR2006-0195 and  FPA 2005/1678.

\end{document}